\documentclass[11pt]{article}

\usepackage[hidelinks,colorlinks=true,linkcolor=blue,citecolor=blue]{hyperref}

\usepackage{amsmath}
\usepackage{graphicx}
\usepackage{enumerate}
\usepackage{natbib}
\usepackage{url} % not crucial - just used below for the URL

\usepackage{xr} % Necessary to reference to Fig and Tab in
\usepackage{amsthm}
\theoremstyle{definition} % Non-italic type theorem
\usepackage{mathtools} %  \overline{text}
\usepackage[usenames,dvipsnames]{color}

\usepackage{enumerate} % Package useful to set symbol in itemize-like
\usepackage[framemethod=TikZ]{mdframed}

\usepackage[labelsep=endash]{caption}
\newcounter{Algocount}
\newcounter{tempFigure}

\usepackage{array}
\newcolumntype{L}{>{\arraybackslash}m{.92\textwidth}}

\usepackage{bbm} % To get the indicator function via \mathbbm{1}

\newcommand{\E}{\mathbb{E}}

\usepackage{amsfonts} % \mathbb{set} where set is R, N, Z, etc for  |R, |N, etc.
\usepackage{eurosym}

\usepackage{epsf}
\usepackage{graphicx}
\usepackage{rotating}
\usepackage{tabularx}

\usepackage{arydshln}

\usepackage[latin1]{inputenc}

\newcommand{\beq}{\begin{equation}}
\newcommand{\eeq}{\end{equation}}

\newcommand{\rit}{{\rm I\!R}}

\newcommand{\ttf}[1]{{\ttfamily{#1}}}

\newcommand{\ee}[1]{\mathrm{e}^{#1}}

\newcommand{\kron}{\otimes}
\newcommand{\diag}{\mathrm{diag}}

\newcommand{\N}[2]{{\cal N}\left(#1,#2\right)}

\newcommand{\G}[2]{{\cal G }\left(#1,#2\right)}
\newcommand{\D}{{{\cal D}}}

\DeclareMathAlphabet\mathbfcal{OMS}{cmsy}{b}{n}

\newcommand{\Zeta}{\mathrm{Z}}

\begin{document}

\bibliographystyle{natbib}

\def\spacingset#1{\renewcommand{\baselinestretch}%
 {#1}\small\normalsize} \spacingset{1}

%%%%%%%%%%%%%%%%%%%%%%%%%%%%%%%%%%%%%%%%%%%%%%%%%%%%%%%%%%%%%%%%%%%%%%%%%%%%%%
\title{\bf
  Penalty parameter selection  and 
  asymmetry corrections to Laplace approximations \\ in Bayesian
  P-splines models
}
% \date{May 11, 2022}
\author{{\Large P}HILIPPE {\Large L}AMBERT$^{*,1,2}$ and
  {\Large O}SWALDO {\Large G}RESSANI$^{3}$,
  \\ [5mm]
  $^1$ Institut de Mathématique,
  Universit\'e de Li\`ege, Belgium\\
  \\ %[3mm]
$^2$ Institut de Statistique, Biostatistique et Sciences Actuarielles
(ISBA), \\
Universit\'e catholique de Louvain, Belgium \\
% Voie du Roman Pays 20, B-1348 Louvain-la-Neuve, Belgium.
\\%[3mm]
$^3$ Interuniversity Institute for Biostatistics and statistical \\ Bioinformatics (I-BioStat), Data Science Institute, \\Hasselt University, Belgium
\\[3mm]
  $^*$ Corresponding author: \ttf{p.lambert@uliege.be}
}

\maketitle

%%% Abstract
%%% -------------------------------------------------------------------------------
\begin{abstract}
  Laplacian-P-splines (LPS) associate the P-splines smoother and the
  Laplace approximation in a unifying framework for fast and flexible
  inference under the Bayesian paradigm. Gaussian Markov field priors
  imposed on penalized latent variables and the Bernstein-von Mises
  theorem typically ensure a razor-sharp accuracy of the Laplace
  approximation to the posterior distribution of these variables. This
  accuracy can be seriously compromised for some unpenalized
  parameters, especially when the information synthesized by the prior
  and the likelihood is sparse. We propose a refined version of the
  LPS methodology by splitting the latent space in two subsets. The
  first set involves latent variables for which the joint posterior
  distribution is approached from a non-Gaussian perspective with an
  approximation scheme that is particularly well tailored to capture
  asymmetric patterns, while the posterior distribution for parameters
  in the complementary latent set undergoes a traditional treatment
  with Laplace approximations. As such, the dichotomization of the
  latent space provides the necessary structure for a separate
  treatment of model parameters, yielding improved estimation accuracy
  as compared to a setting where posterior quantities are uniformly
  handled with Laplace. In addition, the proposed enriched version of
  LPS remains entirely sampling-free, so that it operates at a
  computing speed that is far from reach to any existing Markov chain
  Monte Carlo approach. The methodology is illustrated on the additive
  proportional odds model with an application on ordinal survey data.
\end{abstract}

%%% Key words
%%% -------------------------------------------------------------------------------
\noindent%

%%% Title page
%%% -------------------------------------------------------------------------------
%%% Use command\maketitle to produce the title page.
\maketitle

%%% Main text
%%% ------------------------------------------
\section{Motivation} \label{Intro:Sec}
By publishing his
{\it M\'emoire sur la probabilit\'e des causes par les \'ev\'enements} 
% Memoir on the Probability of the Causes of Events
\citep{Laplace:1774},
% in 1774 \citep{stigler1986a,stigler1986b},
the young French polymath Pierre-Simon de Laplace (1749-1827) seeded
an idea today known as the Laplace approximation. At that time,
Laplace probably could not have imagined that almost two centuries
later, his approximation technique would be resurrected \citep[see
e.g.][]{leonard1982simple, tierney1986accurate, rueinla2009} to play a
pivotal role in modern Bayesian literature. Essentially, the Laplace
approximation is a Gaussian distribution centered around the maximum
\textit{a posteriori} (MAP) of the target distribution with a
variance-covariance matrix that coincides with the inverse of the
negative Hessian of the log-posterior target evaluated at the
MAP. Recently, the ingenuity of Laplace's approximation crossed path
with P-splines, the brainchild of Paul Eilers and Brian Marx
\citep{eilers1996flexible} to inaugurate a new approximate Bayesian
methodology labelled ``Laplacian-P-splines'' (LPS) with promising
applications in survival analysis \citep{gressani2018fast,
  gressani2022laplacian}, generalized additive models
\citep{gressani2021laplace}, nonparametric double additive
location-scale models for censored data \citep{Lambert2021fast} and
infectious disease epidemiology \citep{gressani2021epilps}. The
sampling-free inference scheme delivered by Laplace approximations
combined with the possibility of smoothing different model components
with P-splines in a flexible fashion paves the way for a robust and
much faster alternative to existing simulation-based methods. At the
same time, the LPS toolbox helps P-spline users gain access to the full
potential of Bayesian methods, without having to endure the long and
burdensome CPU- and real-time often required by Markov chain Monte
Carlo (MCMC) samplers.

Although LPS shares some methodological aspects with the popular
integrated nested Laplace approximations (INLA) approach
\citep{rueinla2009}, there are fundamental points of divergence worth
mentioning. First, the tools in INLA and its associated R-INLA
software are originally built to compute approximate posteriors of
univariate latent variables, contrary to LPS that natively delivers
approximations to the (multivariate) joint posterior distribution of
the latent vector. The key benefit of working with an approximate
version of the joint posterior is that pointwise estimators and
credible intervals for subsets of the latent vector (and functions
thereof) can be straightforwardly constructed. Second, by working with
closed-form expressions for the gradients and Hessians involved in the
model, LPS is computationally more efficient than the numerical
differentiation treatment proposed in INLA. Third, while INLA can be
combined with various techniques for smoothing nonlinear model
components, LPS is entirely devoted to P-splines smoothers with the
key advantage of having a full control over the penalization scheme
(as the approximate posterior distribution of the penalty parameter(s)
is analytically available) and in that direction, LPS has a closer
connection to the work of \cite{Wood:Fasiolo:2017}, especially in
the class of (generalized) additive models \citep{Wood:2017}.

The success of Laplace approximations in Bayesian statistics owes much
to a central limit type argument. Under certain regularity conditions,
the Bernstein-von Mises theorem \citep[see e.g.][]{van2000asymptotic}
ensures that posterior distributions in differentiable models converge
to a Gaussian distribution under large samples. In situations
involving small to moderate sample sizes, the asymptotic validity of
the Laplace approximation can become seriously shattered as it does
not account for features involving non-zero skewness (i.e.~lack of
symmetry) \citep{ruli2016improved}. Even under
relatively large samples, the Laplace approximation might fail in
scenarios involving binary data as the latter are poorly informative
for the model parameters and can result in a flat log-likelihood
function, thus complicating inference \citep{ferkingstad2015improving,
  gressani2021laplace}.

Laplacian-P-splines originally belong to the class of latent Gaussian
models, where model parameters are dichotomized between a vector of
latent variables $\boldsymbol{\xi}$ (including penalized B-spline
coefficients, regression coefficients and other parameters of
interest) that are assigned a Gaussian prior and another vector of
hyperparameters $\boldsymbol{\eta}$ that involves nuisance parameters,
such as the smoothing parameter inherent to P-splines, and for which
prior assumptions need not be Gaussian. Combining Bayes' rule and a
simplified Laplace approximation, the conditional posterior
distribution of $\boldsymbol{\xi}$ under the LPS framework is
approximated by a Gaussian distribution denoted by
$\widetilde{p}_G(\boldsymbol{\xi} \vert \widehat{\boldsymbol{\eta}},
\mathcal{D})$, where $\widehat{\boldsymbol{\eta}}$ is a summary
statistic of the posterior hyperparameter vector (e.g. the MAP or
posterior mean/median) and $\mathcal{D}$ denotes the observed
data. Although the latter approximation is typically accurate for
penalized B-spline coefficients, it might be less appropriate for
other candidates in $\boldsymbol{\xi}$ with large prior variance. In
that case, the misfit between the Laplace approximation and a
potentially asymmetric (or heavy-tailed) target posterior distribution
for a parameter can have a detrimental effect on posterior summary
statistics and on any results relying on the generated approximation
for the posterior distribution of the model parameters. This motivates
us to develop an approach that corrects for potential posterior
misfits provided by the Laplace approximation.

\begin{figure}%[h!]
	\centering
	\includegraphics[width=.95\textwidth]{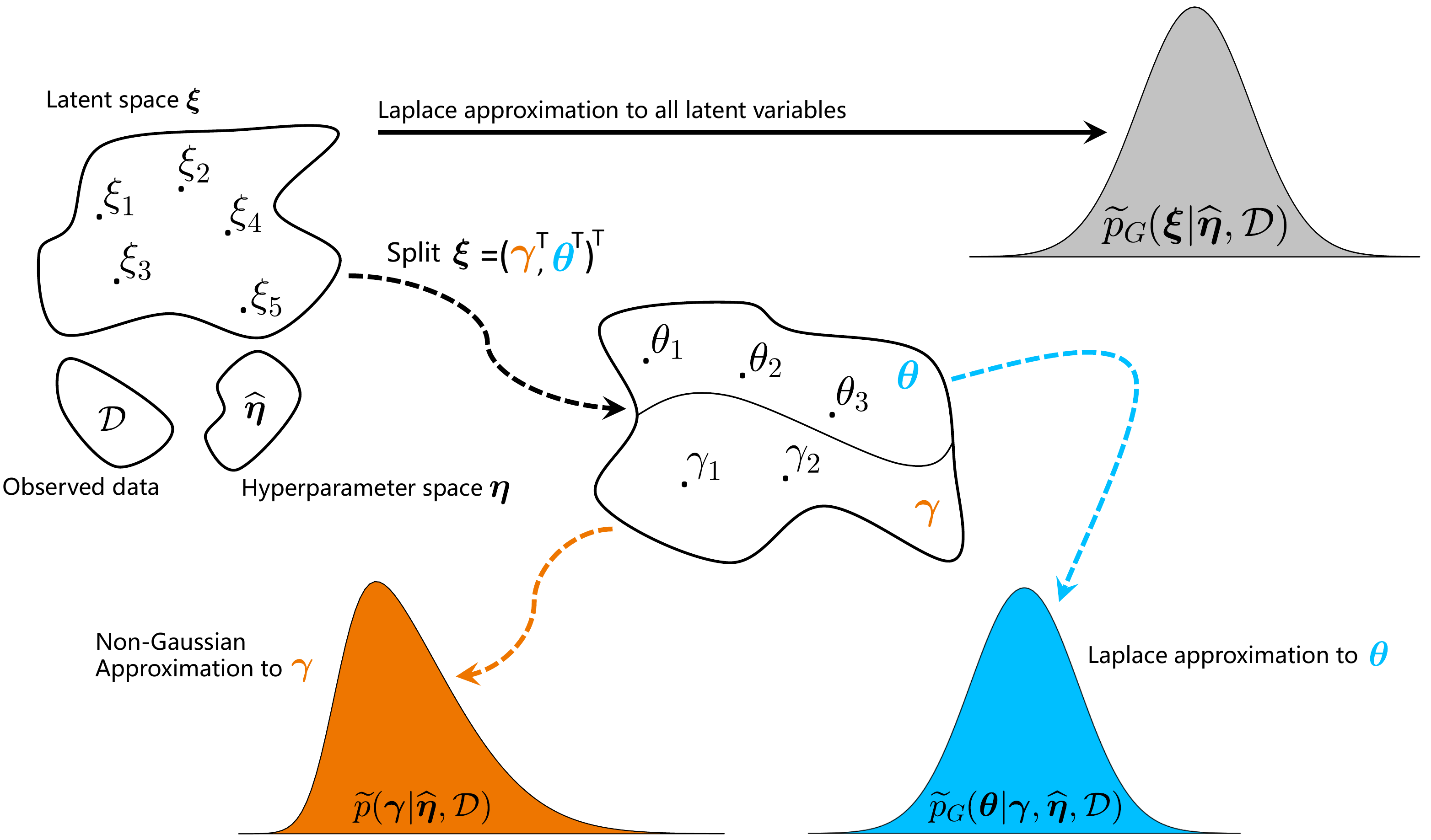}
        \caption{A different
          approximation scheme is proposed for disjoint subsets of the
          latent space. As such, the asymmetric patterns for
          parameters that are suspected to have posterior
          distributions deviating from Gaussianity can be captured
          more accurately.} 
	\label{Fig1}
\end{figure}

A recent technique proposed by \cite{chiuchiolo2022joint} in
the INLA framework consists in using a skew Gaussian copula to correct
for skewness when posterior latent variables have a non-negligible
deviation from Gaussianity. Our proposal in models involving P-splines
consists in splitting the latent parameter space into a set of
parameters $\boldsymbol{\gamma}$ for which the posterior distribution
(conditional on the hyperparameters) is approximated in a non-Gaussian
fashion with an emphasis on capturing asymmetries, and a set of
parameters $\boldsymbol{\theta}$ for which the conditional posterior
is approached with Laplace approximations. Figure~\ref{Fig1}
illustrates the separation of the latent space into two subsets,
i.e.~$\boldsymbol{\xi}=(\boldsymbol{\gamma}^{\top},
\boldsymbol{\theta}^{\top})^{\top}$. The posterior approximation
scheme for $\boldsymbol{\gamma}$ takes asymmetric patterns into
account, while the latent variables in $\boldsymbol{\theta}$ (that
typically involves penalized B-spline coefficients) are approximated
by a Gaussian density. Our refined LPS approach thus allows to obtain
an approximate version of the joint posterior distribution for all the
components in $\boldsymbol{\xi}$ together with a posterior
approximation to the hyperparameter components in $\boldsymbol{\eta}$
without relying on an MCMC sampling scheme. 

%\vspace{0.3cm}

\begin{figure}%[h!]
	\centering
	\includegraphics[width=.95\textwidth]{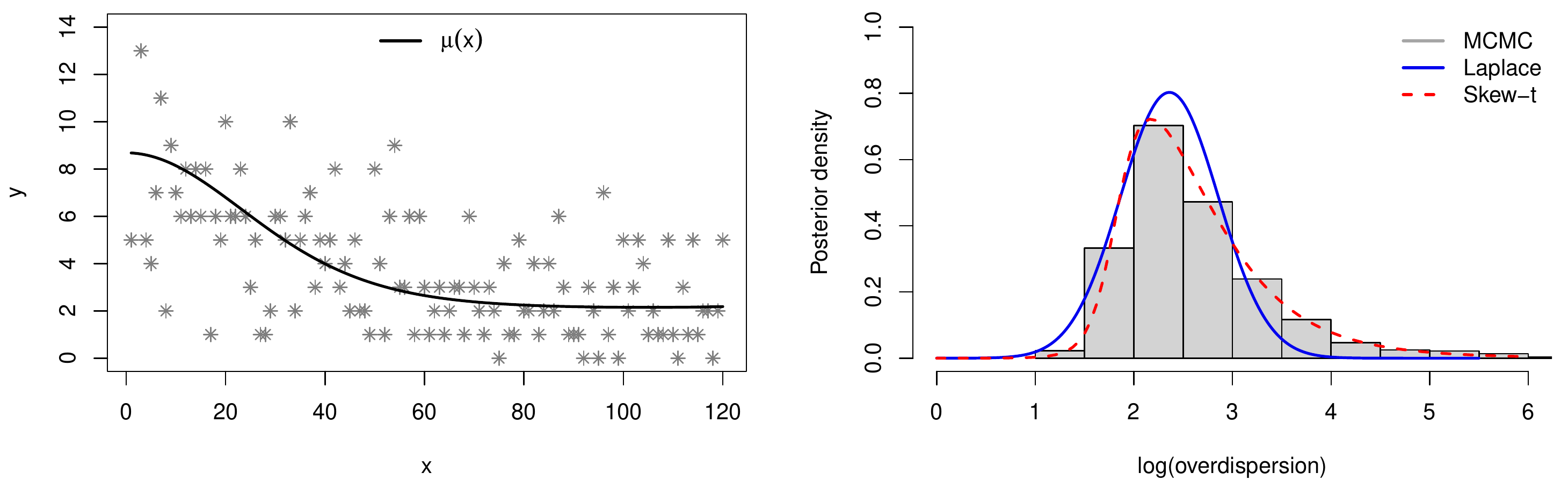}
	\caption{(Left panel) Sample of size $n=120$ generated with a
          nonlinear function $\mu(x)$ in the negative binomial
          model. (Right panel) Laplace approximation misfit to the
          posterior distribution of the log overdispersion
          parameter. Histogram for the posterior of $\log(\gamma)$
          resulting from a long MCMC chain, Laplace approximation
          (solid) and skew-$t$ approximation (dashed).} 
	\label{Fig2}
\end{figure}

A simple motivating example inspired from the infectious disease model
of \cite{gressani2021epilps} helps framing the problem. Let
$\mathcal{D}=\{y_1,\dots,y_n\}$ be an i.i.d. sample of size $n$ from a
negative binomial distribution $\text{NB}(\mu(x), \gamma)$ having a
probability mass function following the parameterization of
\cite{piegorsch1990maximum} with mean $E(y\vert x)=\mu(x)$, variance
$V(y\vert x)=\mu(x)+\mu(x)^2/\gamma$ and overdispersion parameter
$\gamma>0$. We model the mean with P-splines
$\log(\mu(x))=\boldsymbol{\theta}^{\top}\textbf{b}(x)$, where
$\textbf{b}(\cdot)$ is a cubic B-spline basis on the interval $[1,n]$
and $\boldsymbol{\theta}$ is a vector of B-spline coefficients. Note
that $\lim_{\gamma \to +\infty} V(y\vert x)=\mu(x)$, i.e. a Poisson is
obtained as a limiting distribution when the overdispersion parameter
tends to a large number. Following \cite{lang2004bayesian}, a Gaussian
prior is imposed on the vector of B-spline parameters
$(\boldsymbol{\theta}\vert \lambda)\sim
\mathcal{N}_{\text{dim}(\boldsymbol{\theta})}(0,(\lambda P)^{-1})$,
where $P$ is a (full rank) penalty matrix and $\lambda$ is the penalty
parameter to which we assign an uninformative Gamma prior (with mean
$a/b$ and variance $a/b^2$), denoted by
$\lambda \sim \mathcal{G}(a, b)$. To close the Bayesian model, a Gamma
prior with large variance is also imposed on the overdispersion
parameter $\gamma$. The left panel of Figure~\ref{Fig2} shows a data
set of size $n=120$ simulated from the above negative binomial model
with a nonlinear function for $\mu(x)$ and $\gamma=6$. The histogram
for $\log(\gamma)$ on the right panel of Figure~\ref{Fig2} is obtained
from a long MCMC chain with a Metropolis-within-Gibbs algorithm.
There is a serious misfit between the Laplace approximation (solid curve)
and the MCMC output, so that quantities like the posterior standard
deviation or selected posterior quantiles for $\log\gamma$ will be
poorly approximated by Laplace. The dashed curve represents the
skewed distribution that that we propose as an alternative candidate
to the Laplace approximation and that will be thoroughly discussed in
the next section within a Bayesian P-splines context. The fitted
skewed distribution is able to capture the asymmetry that is
apparent in the MCMC sample, improving the precision of posterior
estimates for $\log\gamma$ as compared to Laplace and at a much lower
computational cost as compared to MCMC. As such, the proposed
asymmetry corrections to Laplace approximations will increase the
quality of posterior estimates at virtually no additional
computational cost. The article is organized as follows. Section 2
presents the Bayesian Laplacian-P-splines model and gives a detailed
description of the proposed asymmetric posterior approximation
methodology for non-penalized parameters. In Section 3, we illustrate
the method in an additive proportional odds model for ordinal
data. Finally, Section 4 concludes with a discussion.

%%%%
\section{Laplace approximation and Bayesian P-splines} \label{Methodo:Sec}
\subsection{Model specification} \label{ModelSpec:Sec}
Consider a regression model describing the conditional distribution of
a response $y$ for given covariates $\pmb{x}$. Denote by $\pmb{\xi}$ the
model parameters: it includes the regression and spline parameters,
plus possibly the (log of the) scale and (unconstrained transformed) shape
parameters. Denote by $p(\pmb{\xi}|\pmb{\eta})$ the joint prior
density of $\pmb{\xi}$ conditionally on a vector of hyperparameters
$\pmb{\eta}$. In the context of a P-spline model, the latter might include
$J$ unknown smooth functionals specified as
$f_j(\cdot)=\sum_{\ell=1}^L \theta_{\ell j}b_{j\ell}(\cdot)$
($j=1,\ldots,J$) where 
${\cal B}_j=\{b_{j\ell}(\cdot):\ell=1,\ldots,L\}$ denotes
a B-spline basis associated to equidistant knots spanning the argument
range. Vector $\pmb{\eta}$ would typically include positive roughness penalty
parameters $\lambda_j$ with prior density $p(\lambda_j)$ for the $j$th
functional. The frequentist penalty on changes in differences of
neighbour spline parameters \citep{eilers1996flexible} can be translated
in a Bayesian context using a conditional prior on
$\pmb{\theta}_j=(\theta_{j1},\ldots,\theta_{jL})^\top$
\citep{lang2004bayesian},
$
p(\pmb{\theta}_{j}|\lambda_j) \propto
\exp\left(-{1\over 2}~{\pmb{\theta}_{j}}^\top (\lambda_{j}\mathbf{P})
  \pmb{\theta}_{j} \right)
$,
with $\mathbf{P}=\mathbf{D}_r^\top\mathbf{D}_r$ denoting a penalty
matrix corresponding to a finite difference penalty matrix
$\mathbf{D}_r$ of order $r$. For example, when $r=2$, one has
$\pmb{\theta}_{j}^\top \mathbf{P} \pmb{\theta}_{j}= ||\mathbf{D}_r
\pmb{\theta}_{j}||_2^2 =\sum_{\ell=1}^{L-2}
(\theta_{\ell+2,j}-2\theta_{\ell+1,j}+\theta_{\ell,j})^2$.  The
penalty parameter $\lambda_j>0$ is used to tune the smoothness of the
associated additive term with, at the limit when
$\lambda_j\rightarrow +\infty$, a polynomial of order $r-1$ for
$f_j(\cdot)$.
Different prior distributions could be chosen for
$\lambda_j$ with \citet{BrezgerLang:2006} suggesting to take Gamma priors
$\lambda_j\sim\G{a_j}{b_j}$ (with mean $a_j/b_j$ and variance
$a_j/b_j^2$). A small value for $b_j$ ($=10^{-4}$, say) combined with
$b_j=a_j$ or $a_j=1$ ensures a large prior variance with some more
weight set on small or large values of $\lambda_j$, respectively.
Mixtures of Gamma densities were also investigated in
\citet{JullionLambert:2007}
with $(\lambda|\delta)\sim\G{\nu/2}{\nu\delta/2}$ and
$\delta\sim\G{a_\delta}{b_\delta}$ yielding, in the special case
$a_\delta=b_\delta=.5$ and $\nu=1$, a Beta prime distribution
${\cal B}'(.5,.5)$ for $\lambda$ or a half-Cauchy prior for
$\sqrt{\lambda}$ with   $p(\lambda)\propto
\lambda^{-.5}(1+\lambda)^{-1}$ \citep{LambertBremhorst:2019}.

Penalties can also be combined and sophisticated in multiple ways, see
e.g.~the book by \citet{EilersMarx:2021} for inspiring examples.
More generally, we assume that the joint conditional prior for the
vector $\pmb{\theta}$ stacking all the vectors of penalized B-spline
coefficients in the model can be written as
\begin{align}
  p(\pmb{\theta}|\pmb{\lambda}) \propto
  \exp\left(-{1\over 2}~{\pmb{\theta}}^\top \mathbfcal{P}_\lambda\,{\pmb{\theta}}
  \right)\,,
  \label{Theta:GMRF:Eq}
\end{align}
where $\mathbfcal{P}_\lambda$ is a positive semi-definite matrix.  The
vector of model parameters can be reorganized as follows,
$\pmb{\xi}=(\pmb{\gamma}^\top,\pmb{\theta}^\top)^\top\in\rit^{k_1+k_2}$,
where $\pmb{\gamma}\in\rit^{k_1}$ denotes the vector of non-penalized
parameters. If $\D$ generically denotes the available data and if
$\pmb{\lambda}$ stands for the vector of hyperparameters $\pmb{\eta}$
in the specific context of P-spline models, then the joint posterior
for $\pmb{\xi}$ directly follows from Bayes theorem,
$$
p(\pmb{\xi},\pmb{\lambda}|\D) \propto
\mathcal{L}(\pmb{\xi}|\D)
\, p(\pmb{\gamma})
\,p(\pmb{\theta}|\pmb{\lambda})\,p(\pmb{\lambda}). 
$$
It is typically explored using Markov chain Monte Carlo methods
(MCMC). In this paper, we build up on the methodology described in
\citet{gressani2018fast,gressani2021laplace}
and in \citet{Lambert2021fast} where Laplace
approximations to the conditional posterior of
$(\pmb{\xi}|\pmb{\lambda},\D)$ and an additional approximation to the 
marginal posterior of $(\pmb{\lambda}|\D)$ enable to bypass sampling
algorithms, see Section \ref{PenalySelection:Sec}.

\subsection{Laplace approximation and penalty parameter selection} \label{PenalySelection:Sec}
Assume that closed form expressions can be obtained for the gradient
and Hessian of $\log p(\pmb{\xi}|\pmb{\lambda},\D)$,
\begin{align*}
  \mathbf{U}_{\lambda} = \mathbf{U}_{\lambda}(\pmb{\xi})
  &= {\partial \log p(\pmb{\xi}|\pmb{\lambda},\D) \over \partial
    \pmb{\xi}}
    ~~;~~
    \mathbf{H}_{\lambda} = \mathbf{H}_{\lambda}(\pmb{\xi})
    = {\partial^2 \log p(\pmb{\xi}|\pmb{\lambda},\D) \over
    \partial \pmb{\xi} \partial \pmb{\xi}^\top}~.
\end{align*}
The conditional posterior mode $\hat{\pmb{\xi}}_\lambda$ of
$\pmb{\xi}$ can be quickly obtained using the Newton-Raphson (NR)
algorithm with the substitution,
$ \pmb{\xi} \longleftarrow \pmb{\xi} -
\mathbf{H}_{\lambda}^{-1}\mathbf{U}_{\lambda} $, repeated till
convergence. The Levenberg-Marquardt algorithm
\citep{Marquardt1963,Commenges2006}
%(Marquardt, 1963 ; Commenges {\it et al.}, 2006)
could be preferred if good initial
conditions are not easily found to ensure convergence.  A Laplace
approximation to the conditional posterior distribution of $\pmb{\xi}$
directly follows:
$(\pmb{\xi}|\pmb{\lambda},\D) \stackrel{\cdot}{\sim}
{\cal N}_{k_1+k_2}(\hat{\pmb{\xi}}_\lambda,\Sigma_{\lambda})$ where
$\Sigma_{\lambda}=-\mathbf{H}_{\lambda}^{-1}$.
Thanks to the Gaussian Markov field (GMRF) prior \citep{RueHeld:2005},
$p(\pmb{\theta}|\pmb{\lambda})$, assumed in (\ref{Theta:GMRF:Eq}) for 
the penalized parameters, the Normal approximation to the conditional
posterior of $(\pmb{\theta}|\pmb{\lambda},\D)$ is usually excellent,
see \citet{rueinla2009} for the same argument in latent
Gaussian models. However this might not be true for some non penalized
parameters in $\pmb{\xi}$, especially when the combined information
coming from their prior and the likelihood is sparse, see Section
\ref{AsymmetricPosterior:Sec} for a specific handling.

The preceding Laplace approximation can be used to approximate the
marginal posterior distribution of the penalty 
parameters $\pmb{\lambda}$ with the Normal approximation substituted
in the denominator of the following identity,
$p_\lambda(\pmb{\lambda}|\D) =
p(\pmb{\xi},\pmb{\lambda}|\D)/p(\pmb{\xi}|\pmb{\lambda},\D)
$,
% \begin{align*}
%   p_\lambda(\pmb{\lambda}|\D)
%   & = {p(\pmb{\xi},\pmb{\lambda}|\D) \over
%     p(\pmb{\xi}|\pmb{\lambda},\D)}
% \end{align*}
yielding
\begin{align}
  \widetilde{p}_\lambda(\pmb{\lambda}|\D)
  &= {p(\pmb{\xi},\pmb{\lambda}|\D) \over
    \widetilde{p}_G(\pmb{\xi}|\pmb{\lambda},\D)}
  \propto p(\hat{\pmb{\xi}}_\lambda,\pmb{\lambda}|\D)
    \begin{vmatrix}\widehat{\Sigma}_{\lambda}\end{vmatrix}^{\frac{1}{2}}
  \nonumber \\
  & \propto \underbrace{\mathcal{L}(\hat{\pmb{\xi}}_\lambda|\D)
    \,p(\hat{\pmb{\xi}}_\lambda|\pmb{\lambda},\D)
    \begin{vmatrix}\widehat{\Sigma}_{\lambda}\end{vmatrix}^{\frac{1}{2}}}_{\text{Marginal
    likelihood}}
    \times p(\pmb{\lambda})\,,
    \label{LambdaPosterior:Eq}
\end{align}
see \citet{tierney1986accurate} for the same strategy in the
approximation of a marginal distribution. One might prefer to work
with $\pmb{\upsilon}=\log\pmb{\lambda}$ and its approximate marginal
posterior,
\begin{align}
  \widetilde{p}_\upsilon(\pmb{\upsilon}|\D) =   \tilde{p}_\lambda(\ee{\pmb{\upsilon}}|\D) \,
  \prod_j \ee{\upsilon_j}\,.
  \label{UpsilonPosterior:Eq}
\end{align}
The maximization of (\ref{LambdaPosterior:Eq})
% or of its first factor
or of the marginal likelihood (as with `empirical Bayes' methods)
can be used to select a specific value for
$\pmb{\lambda}$. Alternatively, it could be made 
from the log-penalty using (\ref{UpsilonPosterior:Eq}), yielding
larger penalty values when the selection is made from the marginal
posterior instead of the (parametrization invariant) marginal likelihood.

\subsection{Asymmetric posterior for non-penalized parameters}
\label{AsymmetricPosterior:Sec}
Assume that $\pmb{\xi} = (\pmb{\gamma}^\top,\pmb{\theta}^\top)^\top\in\rit^{k_1+k_2}$ with
$\pmb{\gamma}\in\rit^{k_1}$ suspected to have a non-symmetric marginal posterior
distribution. Let
$\hat{\pmb{\xi}}_\lambda = (\hat{\pmb{\gamma}}_\lambda^\top,
\hat{\pmb{\theta}}_\lambda^\top)^\top$ denote the posterior mode of
$p(\pmb{\xi}|\pmb{\lambda},\D)$ and
$\widehat{\Sigma}_\lambda^{-1}$ the observed information matrix structured
in blocks as follows,
\begin{align*}
\widehat{\Sigma}_\lambda =
  \begin{bmatrix}
    \widehat{\Sigma}^{\gamma\gamma}_\lambda & \widehat{\Sigma}^{\gamma\theta}_\lambda\\
    \widehat{\Sigma}^{\theta\gamma}_\lambda & \widehat{\Sigma}^{\theta\theta}_\lambda
  \end{bmatrix}~.
\end{align*}
The conditional posterior of
$(\pmb{\theta}|\pmb{\gamma},\pmb{\lambda},\D)$ has an approximate
Normal distribution resulting from the the GMRF prior in
(\ref{Theta:GMRF:Eq}) for $(\pmb{\theta}|\pmb{\lambda})$.
One has
$$
(\pmb{\theta}|\pmb{\gamma},\pmb{\lambda},\D)
\stackrel{\cdot}{\sim}
{\cal N}_{k_2}\left({\E(\pmb{\theta}|\pmb{\gamma},\pmb{\lambda},\D)},
  {\widehat{\Sigma}^{\theta|\gamma}_\lambda}\right)~,
$$
where
\begin{align}
  \begin{split}
  &\E(\pmb{\theta}|\pmb{\gamma},\pmb{\lambda},\D)
    =  \hat{\theta}_\lambda + \widehat{\Sigma}^{\theta\gamma}_\lambda
    \left(\widehat{\Sigma}^{\gamma\gamma}_\lambda\right)^{-1}(\pmb{\gamma}-\hat{\pmb{\gamma}}_\lambda)
  ~;\\
  &\widehat{\Sigma}^{\theta|\gamma}_\lambda
    = \widehat{\Sigma}^{\theta\theta}_\lambda
    - \widehat{\Sigma}^{\theta\gamma}_\lambda
    \left(\widehat{\Sigma}^{\gamma\gamma}_\lambda\right)^{-1}
    \widehat{\Sigma}^{\gamma\theta}_\lambda~.
  \end{split}
    \label{Moments:ThetaGivenGamma:Eq}
\end{align}
Hence, starting from the following identity,
$  p_\gamma(\pmb{\gamma}|\pmb{\lambda},\D)
  = {p(\pmb{\gamma},\pmb{\theta}|\pmb{\lambda},\D)
  / p(\pmb{\theta}|\pmb{\gamma},\pmb{\lambda},\D)}
$, one gets the approximation
\begin{align}
  p_\gamma(\pmb{\gamma}|\pmb{\lambda},\D)
  &\approx {p(\pmb{\gamma},\pmb{\theta}|\pmb{\lambda},\D)
    \over \widetilde{p}_G(\pmb{\theta}|\pmb{\gamma},\pmb{\lambda},\D)}
  % &= (2\pi)^{k_2}
  \propto  p\left(\pmb{\gamma},\E(\pmb{\theta}|\pmb{\gamma},\pmb{\lambda},\D)\big|\pmb{\lambda},\D\right)
    \begin{vmatrix}
      \widehat{\Sigma}^{\theta|\gamma}_\lambda
    \end{vmatrix}^{\frac{1}{2}}~,  \label{MarginalGamma:Eq}
\end{align}
see Eq.\,(2) in \citet{tierney1989fully} for a similar expression.
We propose to reparametrize $\pmb{\gamma}$ by projecting it
on the eigenvectors of the singular value decomposition (SVD) of
$\widehat{\Sigma}^{\gamma\gamma}_\lambda =
\mathbf{V}\Zeta\mathbf{V}^\top$
where
$\mathbf{V}=[\mathbf{v}_1\ldots \mathbf{v}_{k_1}]$
denotes the matrix of orthonormal eigenvectors, $\pmb{\zeta}$ the
eigenvalues and $\Zeta=\diag(\pmb{\zeta})$. It yields
$\tilde{\pmb{\gamma}}=\Zeta^{-\frac{1}{2}}\mathbf{V}^\top(\pmb{\gamma}-\hat{\pmb{\gamma}}_\lambda)$ and
$\pmb{\gamma}=\hat{\pmb{\gamma}}_\lambda+\mathbf{V}\Zeta^{\frac{1}{2}}\tilde{\pmb{\gamma}}$,
with
\begin{align}
  p_{\tilde\gamma}(\tilde{\pmb{\gamma}}|\pmb{\lambda},\D)
  \propto
  p_\gamma(\hat{\pmb{\gamma}}_\lambda+\mathbf{V}\Zeta^{\frac{1}{2}}\tilde{\pmb{\gamma}}|\pmb{\lambda},\D).
  \label{MarginalGammaTilde:Eq}
\end{align}
The posterior dependence between the components of $\tilde{\pmb{\gamma}}$
is expected to be milder than under the original $\pmb{\gamma}$
parametrization. Therefore, conditionally on $\pmb{\lambda}$, we propose to approximate the joint posterior density
of $(\tilde{\pmb{\gamma}}|\pmb{\lambda},\D)$ by the product of the
marginal densities of its components:
\begin{align}
  p_{\tilde\gamma}(\tilde{\pmb{\gamma}}|\pmb{\lambda},\D)
  \approx
     \prod_{s=1}^{k_1}
  p_{\tilde{\gamma}_s}(\tilde{\gamma}_s|\pmb{\lambda},\D).
\label{ApproxMarginalGammaTilde:Eq1}
\end{align}
Under that working independence hypothesis, each univariate marginal in the
product in (\ref{ApproxMarginalGammaTilde:Eq1}) is equal to its
conditional with the other components 
set equal to an arbitrary value. Combined with
(\ref{MarginalGammaTilde:Eq}), it implies that
\begin{align}
  p_{\tilde{\gamma}_s}(\tilde{\gamma}_s|\pmb{\lambda},\D)
  &=  p_{\tilde{\gamma}_s|\tilde{\gamma}_{-s}}(\tilde{\gamma}_s|\tilde{\pmb{\gamma}}_{-s}=0,\pmb{\lambda},\D)
  \nonumber \\    
  &\propto 
    p_{\tilde{\gamma}}(\tilde{\gamma}_s\mathbf{e}_s|\pmb{\lambda},\D)
   \nonumber\\
 &
    \propto 
  p_{\gamma}(\hat{\pmb{\gamma}}_\lambda+\tilde{\gamma}_s\,\sqrt{\zeta_s}\mathbf{v}_s|\pmb{\lambda},\D) ~,
    \label{ApproxMarginalGammaTilde:Eq2}
\end{align}
where $\mathbf{e}_s$ denotes the $s$th unit vector in $\rit^{k_1}$
such that $[\mathbf{e}_s]_k=\delta_{ks}$. The univariate marginal
posterior density in (\ref{ApproxMarginalGammaTilde:Eq2}) can be
evaluated for any value of $\tilde{\gamma}_s$ 
using (\ref{MarginalGamma:Eq}). We suggest to approximate it 
using a skew-$t$ (ST) or a skew-normal (SN) density. Here, we provide
details for the SN density, the generalization to the ST being
straightforward. By definition, 
$X\sim\mathrm{SN}(\psi,\omega^2,\alpha)$ if $X\in\rit$ and has density
$$
\varphi(x|\psi,\omega^2,\alpha) =
{2\over\omega} \phi\left({x-\psi\over\omega}\right)
\Phi\left(\alpha{x-\psi\over\omega}\right)~,
$$
with location parameter $\psi$, scale parameter $\omega>0$ and slant
(or skewness) parameter $\alpha\in\rit$,
see e.g.~\citet{AzzaliniCapitanio:2014} for more details on its definition
and properties. Such an approximation can be obtained by matching the
first three central moments of the target variable with those of the
approximating SN distribution, yielding, after
matching the mean with the posterior mode of the target distribution,
$(\tilde{\gamma}_s|\pmb{\lambda},\D) \stackrel{\cdot}{\sim}
\mathrm{SN}(\tilde\psi_s,\tilde\omega_s^2,\tilde\alpha_s)$.
Substituting these SN densities in (\ref{ApproxMarginalGammaTilde:Eq1})
provides an approximation to the joint posterior of
$(\tilde{\pmb{\gamma}}|\pmb{\lambda},\D)$ and an efficient method to
sample from it using the independence of its components.
An analytic form for the joint posterior density of
$(\pmb{\gamma}|\pmb{\lambda},\D)$ can be obtained by
combining (\ref{ApproxMarginalGammaTilde:Eq1}) with 
$\tilde{\pmb{\gamma}}
=\Zeta^{-\frac{1}{2}}\mathbf{V}^\top(\pmb{\gamma}-\hat{\pmb{\gamma}}_\lambda)$,
yielding
\begin{align*}
  p_\gamma(\pmb{\gamma}|\pmb{\lambda},\D) &=
  \prod_{s=1}^{k_1} {1\over\sqrt{\zeta_s}}\, \varphi(\tilde{\gamma}_s|\tilde{\psi}_s,\tilde{\omega}_s^2,\tilde{\alpha}_s).
\end{align*}
An approximation to the marginal distribution of its $s$th component
$(\gamma_s|\pmb{\lambda},\D)$ can be obtained using fast Monte Carlo techniques.
Indeed, an approximate large random sample from
$(\tilde{\pmb{\gamma}}|\pmb{\lambda},\D)$ can first be generated by 
sampling its independent skew-normal components
$\mathrm{SN}(\tilde\psi_s,\tilde\omega_s^2,\tilde\alpha_s)$, yielding 
$\{\tilde{\pmb{\gamma}}^{(m)}:m=1,\ldots,M\}$.
The associated random sample for $({\pmb{\gamma}}|\pmb{\lambda},\D)$ is given by
$\{\pmb{\gamma}^{(m)}=\hat{\pmb{\gamma}}_\lambda+\mathbf{V}\Zeta^{\frac{1}{2}}\tilde{\pmb{\gamma}}^{(m)}:m=1,\ldots,M\}$.
Then, a skew-normal approximation to the marginal posterior of
$\gamma_s$ can be fitted to $\{\gamma_s^{(m)}:m=1,\ldots,M\}$,
yielding
$(\gamma_s|\pmb{\lambda},\D) \stackrel{\cdot}{\sim}
\mathrm{SN}(\psi_s,\omega_s^2,\alpha_s)$. Point estimates or credible
regions for $\gamma_s$ can be computed from it.

These different steps provide a convenient approximation to the joint
posterior of the model parameters. Indeed, based one the factorization
of the joint posterior density,
$$
  p(\pmb{\xi},\pmb{\lambda}|\D)
  = p(\pmb{\theta}|\pmb{\gamma},\pmb{\lambda},\D)
    \,p(\pmb{\gamma}|\pmb{\lambda},\D)
    \,p(\pmb{\lambda}|\D)~,
$$    
one has the following stochastic representation for $(\pmb{\xi},\pmb{\lambda}|\D)$,
\begin{align}
  (\pmb{\xi},\pmb{\lambda}|\D)
  &\stackrel{\cdot}{\sim}
    {\cal N}_{k_2}\left(\E(\pmb{\theta}|\pmb{\gamma},\pmb{\lambda},\D),
                           {\widehat{\Sigma}^{\theta|\gamma}_\lambda}\right)
    \times
    \prod_{s=1}^{k_1}
    \text{SN}(\tilde{\gamma}_s|\tilde{\psi}_s,\tilde{\omega}_s^2,\tilde{\alpha}_s)
    \times
    (\pmb{\lambda}|\D)~,
    \label{JointPosteriorApprox:Eq}
\end{align}
with moments in the first factor given in
(\ref{Moments:ThetaGivenGamma:Eq}). 
It can be used to generate an arbitrarily large number of independent
copies from the joint posterior incomparably faster than with
MCMC. This is for example particularly useful to make inference on
complicated functions of the model parameters or to visualize the
posterior predictive distribution of the response for given
covariates values.

%%%%
\section{Illustration}
\label{PropOddsModel:Sec}

% %% 
\subsection{The additive proportional odds model for ordinal  data} 
Denote by ${\cal E}_m=\{1,\ldots,m\}$ the set containing the first $m$ positive
integers. Assume that $n$ independent units are observed with data
$\D=\{(y_i,\mathbf{x}_i):i=1,\ldots,n\}$ where
$y$ is an ordinal random variable taking values in 
${\cal E}_R$ and $\mathbf{x}$ a vector of covariates.
The proportional odds (PO) model is a popular choice when the response is
ordinal \citep{Agresti:2010}. It assumes that
\begin{align*}
  & \mathrm{logit}[P(Y \leq r|\mathbf{x})]
    =
    \eta_{r} = \gamma_r + \mathbf{x}^\top\pmb{\theta}
    ~~~~(r \in {\cal E}_{R-1})
\end{align*}
with a specific intercept $\gamma_r$ for each cumulative logit,
but a shared vector of regression parameters
$\pmb{\theta}$. Consequently,
\begin{align*}
 \log {\Pr(Y \leq r|\mathbf{x}_1) / \Pr(Y > r|\mathbf{x}_1)
  \over
  \Pr(Y \leq r|\mathbf{x}_2) / \Pr(Y > r|\mathbf{x}_2)
  }
  =\pmb{\theta}(\mathbf{x}_1-\mathbf{x}_2)
\end{align*}
is independent of $r$ and provides a clear interpretation to the
regression parameters $\pmb{\theta}$ with a change $\theta_k$ in the
log-odds of $Y$ taking values in the lower scale of the ordinal scale
for every unit increase in the $k$th component of $\mathbf{x}$.
Let $F_{ir}=P(Y_i\leq r|\mathbf{x}_i) = \ee{\eta_{ir}}/(1+\ee{\eta_{ir}})$
for $r$ in ${\cal E}_{R-1}$ and let $F_{i0}=0$, $F_{iR}=1$.
Then, $\pi_{ir}=P(Y_i=r|\mathbf{x}_i)=F_{ir}-F_{i,r-1}$
for $r$ in ${\cal E}_{R}$. Let
$\pmb{\xi}=(\pmb{\gamma}^\top,\pmb{\theta}^\top)^\top$ and assume a prior of the
following form,
\begin{align}
  p(\pmb{\xi}|\pmb{\lambda}) \propto
  \exp\left(-{1\over 2}~({\pmb{\xi}}-\mathbf{e})^\top \mathbf{K}_\lambda
  ({\pmb{\xi}}-\mathbf{e})
  \right)~,
  \label{xi:GMRF:Eq}
\end{align}
conditionally on a vector of parameters $\pmb{\lambda}$ and for a
positive semi-definite matrix $\mathbf{K}_\lambda$.
The log-likelihood takes a simple form,
$\ell(\pmb{\xi}|\D)=\sum_{i=1}^n \log \pi_{iy_i}(\pmb{\xi})$,
with the resulting conditional posterior for $\pmb{\xi}$,
\begin{align}
\log p(\pmb{\xi}|\pmb{\lambda},\D)
  = \ell(\pmb{\xi}|\D)
-{1\over 2}~({\pmb{\xi}}-\mathbf{e})^\top \mathbf{K}_\lambda
  ({\pmb{\xi}}-\mathbf{e}).
  \label{xi:CondPost:Eq}
\end{align}
Explicit analytical forms can be obtained for the associated gradient and
Hessian matrix, see Appendix \ref{Appendix:A}.
Assume now for simplicity a model with $J$ continuous covariates
$x_1,\ldots,x_J$ with smooth additive terms $f_j(x_j)$ ($j=1,\ldots,J$)
describing their effects on the conditional log-odds,
\begin{align*}
  & \mathrm{logit}[P(Y \leq r|\mathbf{x})]
    =
    \eta_{r} = \gamma_r + f_1(x_1) +\ldots +f_J(x_J)~.
\end{align*}
Following \citet{eilers1996flexible}, consider now a basis of $(L+1)$ cubic B-splines
$\{s^*_{j\ell}(\cdot)\}_{\ell=1}^{L+1}$ associated to a generous number
of equally spaced knots on the range $(x_j^{\min},x_j^{\max})$ of
values for $x_j$ \citep{MarxEilers1998}.  They are recentered for
identification purposes in the additive model using
$s_{j\ell}(\cdot)= s^*_{j\ell}(\cdot) - {1\over
  x_j^{\max}-x_j^{\min}}\int_{x_j^{\min}}^{x_j^{\max}}s^*_{j\ell}(u)du~(\ell=1,\ldots,L)$.
Then, the additive terms in the conditional model can be approximated
using linear combinations of these recentered B-splines,
$f_j(x_j)=\sum_{\ell=1}^L {s_{j\ell}}(x_{ij})\theta_{\ell j}$.  In a
Bayesian framework, as reminded in Section \ref{ModelSpec:Sec},
smoothness can be forced on these additive terms 
by taking GMRF priors for the vectors of spline coefficients 
\begin{align*}
p(\pmb{\theta}_{j}|\lambda_j) \propto
\exp\left(-{1\over 2}~{\pmb{\theta}_{j}}^\top (\lambda_{j}\mathbf{P}) \pmb{\theta}_{j} \right)~,
\end{align*}
where $\mathbf{P}$ stands for the penalty matrix.
A multivariate Normal prior could be taken
for $\pmb{\gamma}$ to complete the model specification,
$\pmb{\gamma}\sim \N{\tilde{\mathbf{e}}}{{(\mathbf{Q})}^{-1}}$.  Under
the general formulation in (\ref{xi:GMRF:Eq}), one has
$\pmb{\xi}=(\pmb{\gamma}^\top,\pmb{\theta}^\top)^\top$,
$\pmb{\theta}=(\pmb{\theta}_1^\top,\ldots,\pmb{\theta}_J^\top)^\top$,
$\mathbf{e}=(\tilde{\mathbf{e}}^\top,\mathbf{0}_{JL}^\top)^\top$, and a
block-diagonal penalty matrix
$\mathbf{K}_\lambda=
\diag\big(\mathbf{Q},{\mathbfcal{P}}_\lambda\big)$ where
${\mathbfcal{P}_\lambda}=\pmb{\Lambda} \kron \mathbf{P}$ with
$[\pmb{\Lambda}]_{jj'}=\delta_{jj'}\lambda_j$. The conditional
posterior for $\pmb{\xi}$ is given by (\ref{xi:CondPost:Eq}).  With a
Gamma prior for the penalty parameters, $\lambda_j\sim\G{a}{b}$, one has
$(\lambda_j|\pmb{\xi},\D)\sim
\G{a+.5\,\rho(\mathbf{P})}{b+\pmb{\theta}_j^\top\mathbf{P}\pmb{\theta}_j}$.
Starting from these conditional posterior distributions, a
Metropolis-within-Gibbs algorithm can be set up to generate a random
sample from the joint posterior for $(\pmb{\xi},\pmb{\lambda})$, with
Gibbs steps for the penalty parameters $\pmb{\lambda}$ and Metropolis
steps for the regression and splines parameters
$\pmb{\xi}$. Alternatively, proposals for $\pmb{\xi}$ could be made
using the modified Langevin
\citep{RobertsTweedie:1996,LambertEilers:2009}
or the Metropolis-Hastings algorithm with proposals
based on the local topological information provided by the explicit
analytic forms for the gradient and Hessian matrix \citep{Gamerman:1997}.
Such a sampling approach based on MCMC will be compared to the
strategy proposed in Section \ref{Methodo:Sec}.

\subsection{Application on survey data}
\begin{figure}\centering
  \begin{tabular}{cc}
    \includegraphics[width=6cm]{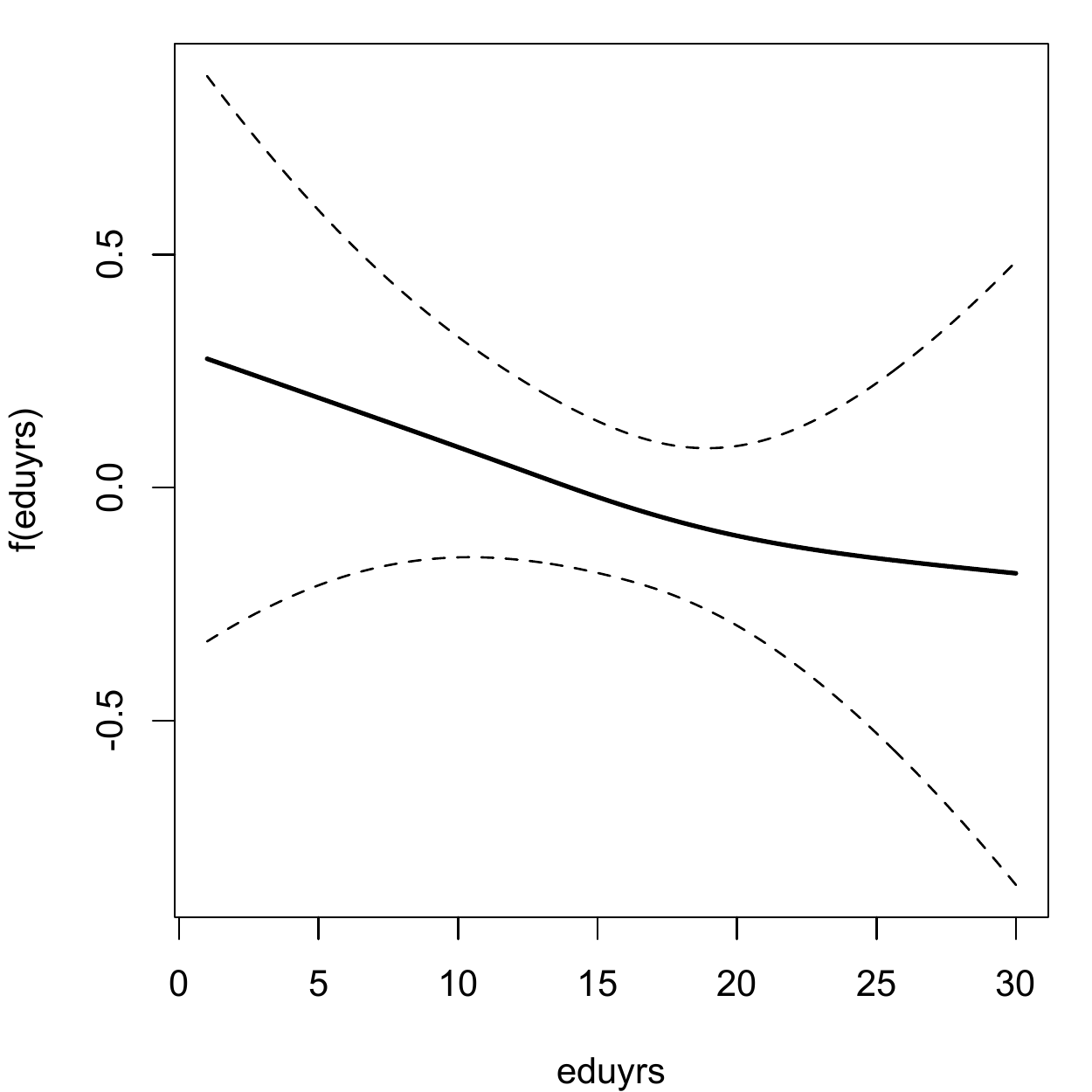} &
    \includegraphics[width=6cm]{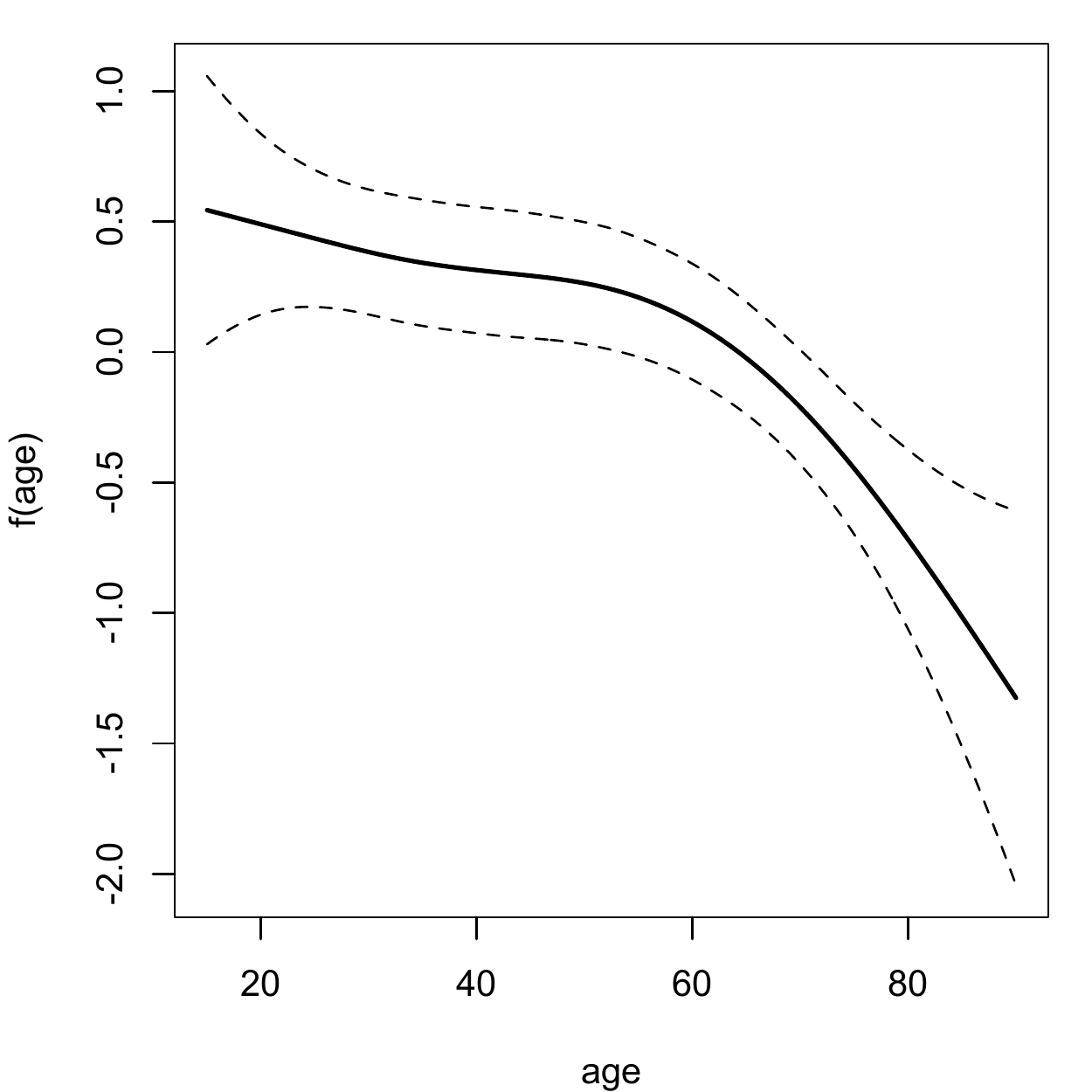}  
  \end{tabular}
\caption{ESS dataset: fitted additive terms for \ttf{eduyrs} and
  \ttf{age} with pointwise 95\% credible intervals.}
 \label{ESS2018:Fig1}
\end{figure}

\begin{figure}\centering
  \includegraphics[width=.49\textwidth]{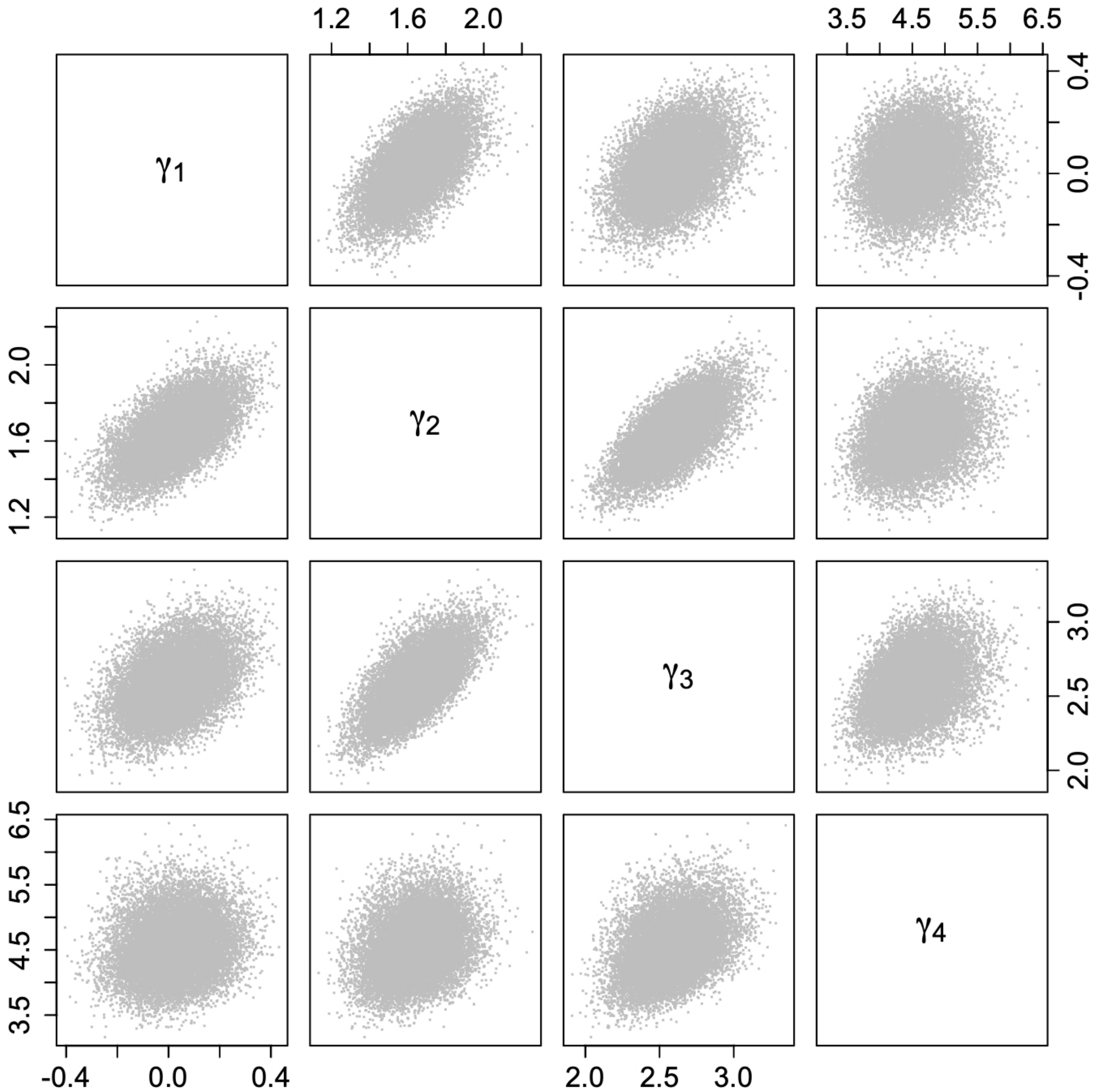}
  \includegraphics[width=.49\textwidth]{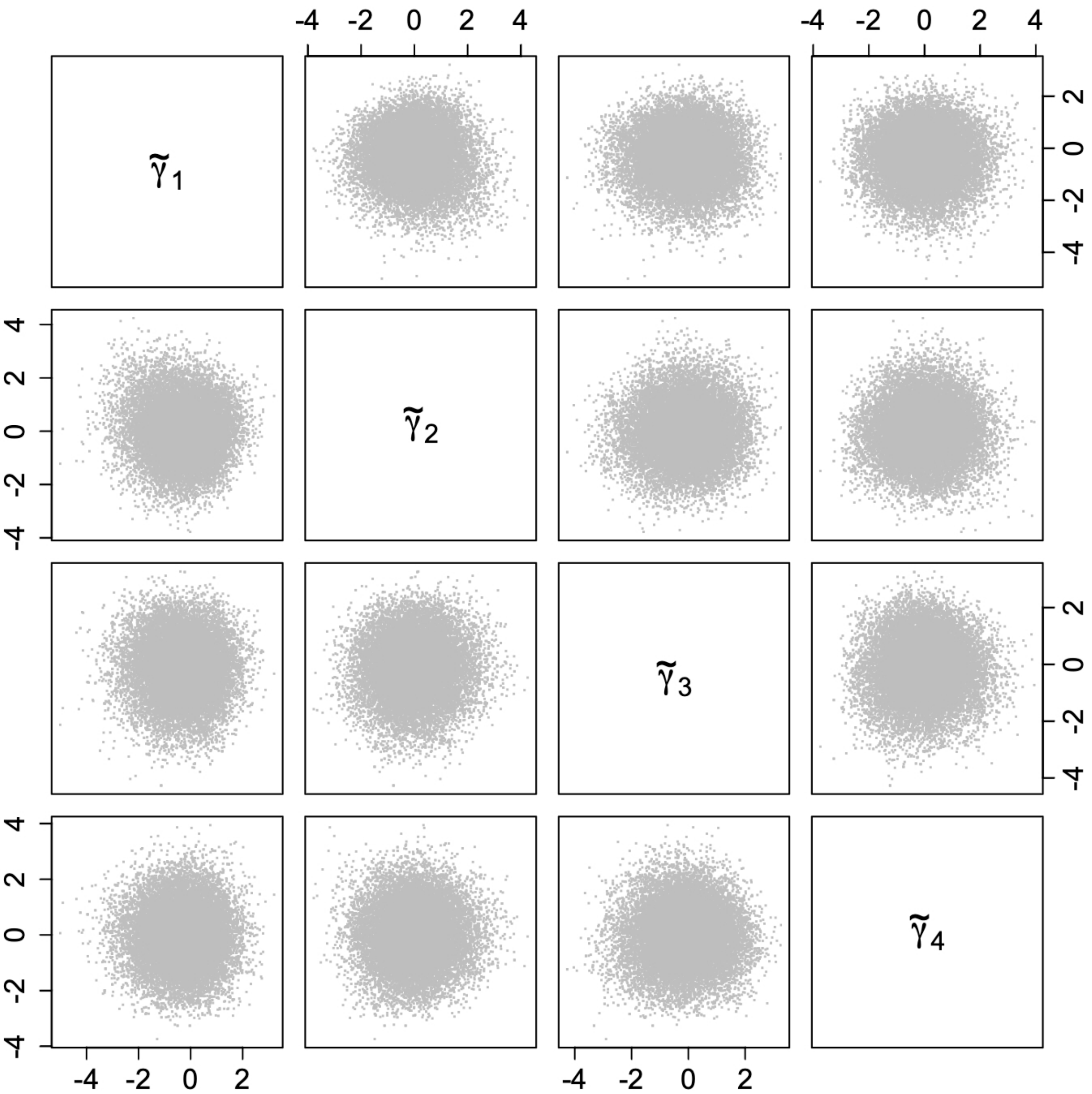}
\caption{ESS dataset: scatterplots of the MCMC samples for 
  $(\pmb{\gamma}|{\pmb{\lambda}},\D)$ and
  $(\tilde{\pmb{\gamma}}|{\pmb{\lambda}},\D)$ 
  when $\pmb{\lambda}=\hat{\pmb{\lambda}}$.}
 \label{ESS2018:MCMCScatterPlots:Fig}
\end{figure}

\begin{figure}\centering
    \includegraphics[width=\textwidth]{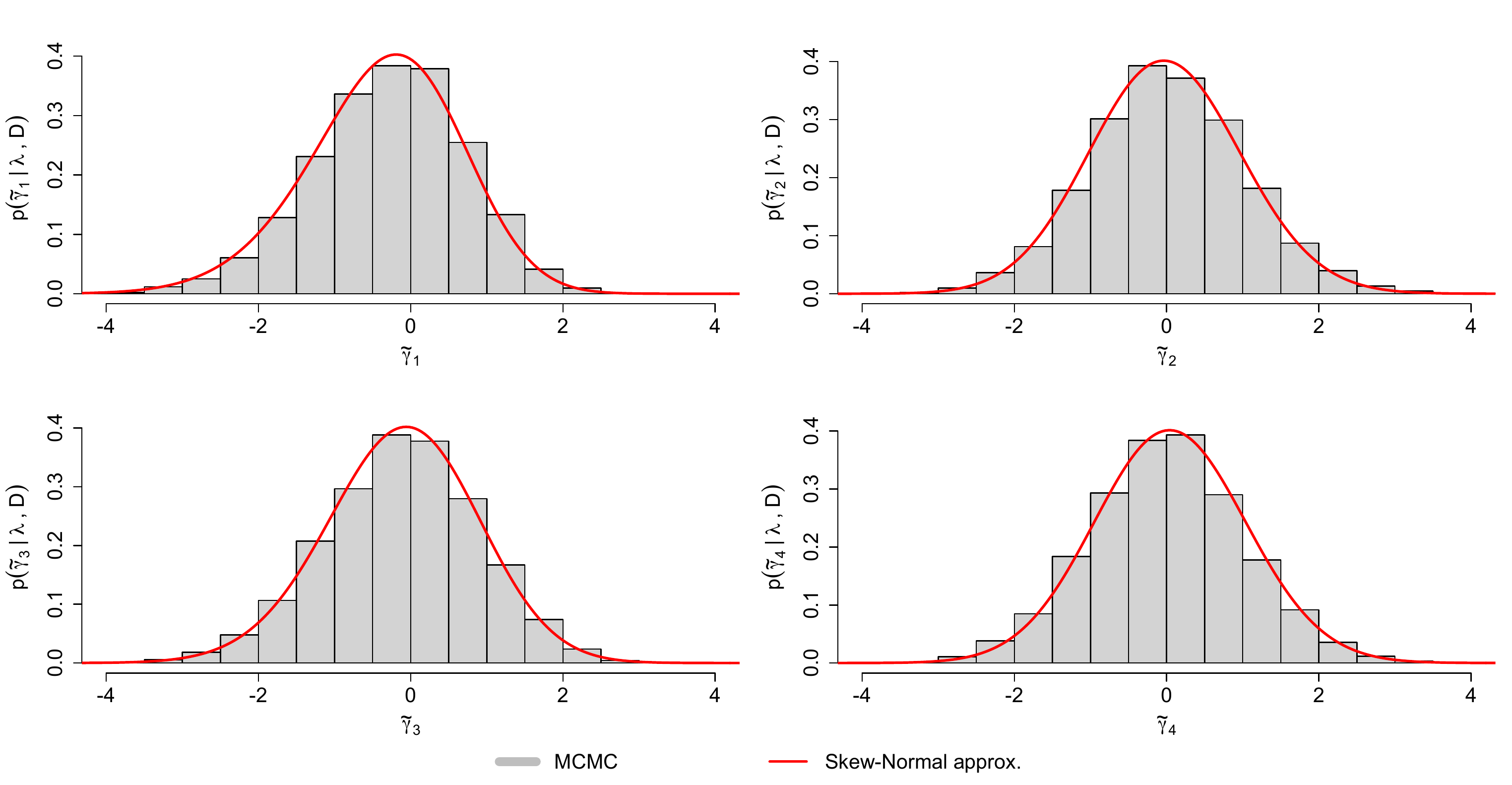}
\caption{ESS dataset: approximated marginal posterior density for
  $(\tilde{\pmb{\gamma}}|{\pmb{\lambda}},\D)$ compared to MCMC
  samples when $\pmb{\lambda}=\hat{\pmb{\lambda}}$.}
 \label{ESS2018:GamtildePosterior:Fig}
\end{figure}

\begin{figure}\centering
    \includegraphics[width=\textwidth]{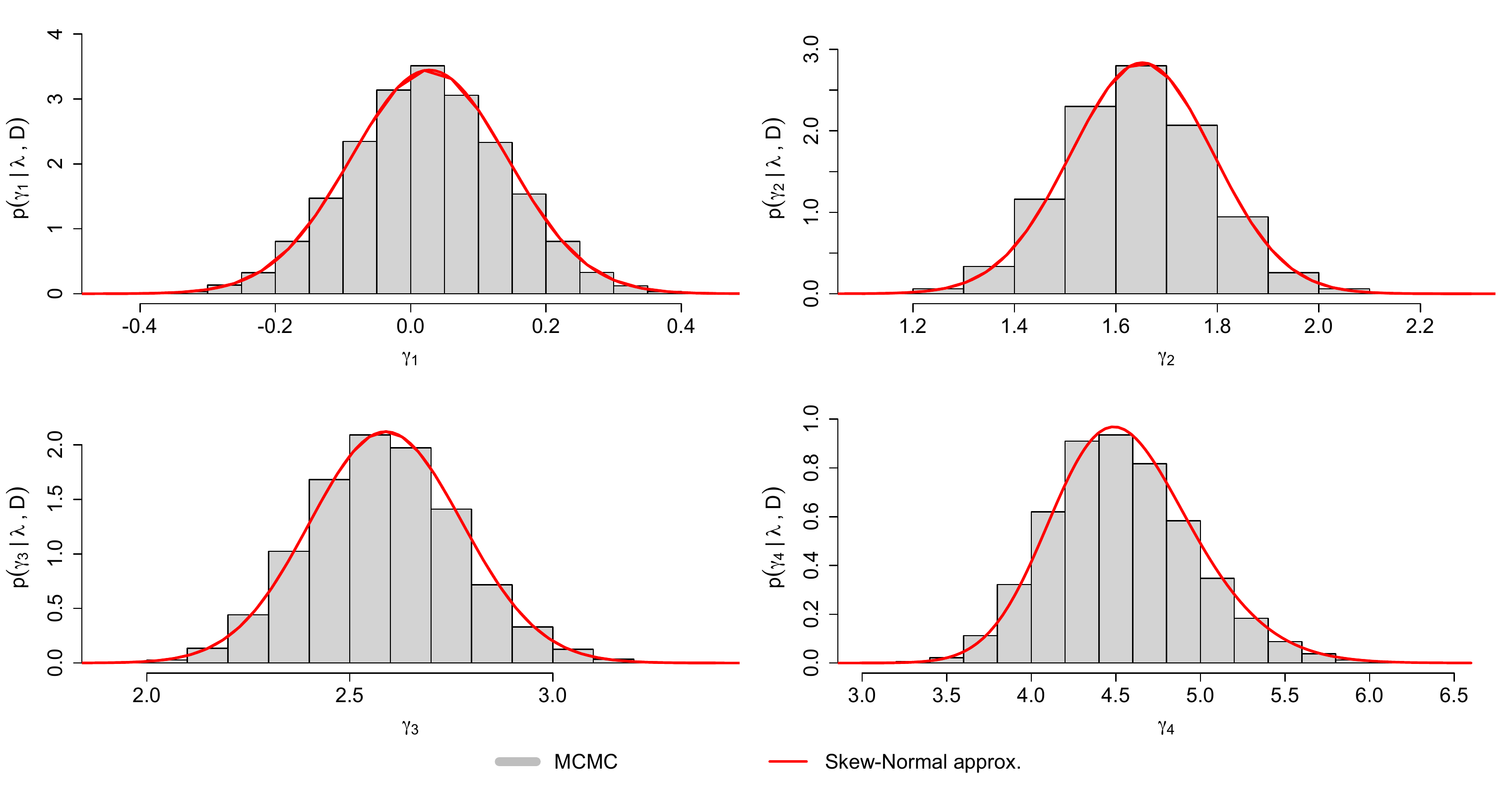}
\caption{ESS dataset: approximated marginal posterior density for 
  $(\pmb{\gamma}|{\pmb{\lambda}},\D)$ compared to MCMC samples
  when $\pmb{\lambda}=\hat{\pmb{\lambda}}$.}
 \label{ESS2018:GammaPosterior:Fig}
\end{figure}

Consider now an illustration of the proposed methodology on data
coming from the European Social Survey
\citep{ESS2018} with a
specific focus on the French speaking respondents from Wallonia, one
of the three regions in Belgium. Each of the participants (aged at
least 15) was asked to react to the following statement, {\it Gay men
  and lesbians should be free to live their own life as they wish},
with a positioning on a Likert scale going from 1 (={\it Agree
  strongly}) to 5 (={\it Disagree strongly}), with 3 labelled as {\it
  Neither agree nor disagree}
(with relative frequencies 1: 54.9\% ; 2: 30.4\% ; 3: 8.2\% ; 4: 5.4\%
; 5: 1.1\%). That ordinal 
response effectively 
recorded on $n=552$ respondents was analyzed
using the proportional odds model described above with the number of completed
years of education ($14.1\pm 4.4$ years) and age ($47.3\pm 18.5$
years) entering as additive terms with $L=10$ recentered B-splines
spanning each covariate range. Starting from Gamma priors,
$\lambda_j\sim\G{1}{10^{-4}}$ ($j=1,2$), the penalty parameters
$\lambda_1$ and $\lambda_2$
associated to $f_1(\text{eduyrs})$ and $f_2(\text{age})$, respectively,
were selected by maximizing $p(\pmb{\lambda}|\D)$ in
(\ref{LambdaPosterior:Eq}) using the Levenberg-Marquardt algorithm, yielding
$\hat\lambda_1=191.8$ (e.d.f.=1.24), $\hat\lambda_2=18.4$ (e.d.f.=2.55),
the value in brackets standing for the effective degrees of freedom.
Alternatively the maximization of the marginal likelihood in
(\ref{LambdaPosterior:Eq}) would yield
a very large value for $\hat\lambda_1$ (e.d.f.=1.00), suggesting
linearity for $f_1(\text{eduyrs})$, and $\hat\lambda_2=18.5$ (e.d.f.=2.55)
practically unchanged. 
The fitted additive terms are visible on Fig.\,\ref{ESS2018:Fig1} with
their pointwise 95\% credible intervals, suggesting a statistically
non-significant effect of \ttf{eduyrs}, but a tolerant perception
of homosexuality tending to decrease with \ttf{age}, with a marked
change in attitude revealed beyond age 60. 
To compare the merits of our methodological proposal, a MCMC algorithm
was run to explore $p(\pmb{\xi}|\hat{\pmb{\lambda}},\D)$, the
generated samples and their properties being compared to the
analytical approximations suggested in Section
\ref{AsymmetricPosterior:Sec}.  The estimated additive terms and their
95\% credible intervals based on MCMC are practically identical to our
estimates in Fig.\,\ref{ESS2018:Fig1}, confirming the excellent
quality of the Laplace approximation to the conditional posterior
distribution of $(\pmb{\theta}|\pmb{\lambda},\D)$ underlying our
calculations. Let us now focus on the non-penalized regression
parameter in $\pmb{\gamma}$ standing for the four intercepts in the
proportional odds model. The scatterplot of the MCMC sample
$\{\pmb{\gamma}^{(m)}:m=1,\ldots,M\}$ generated from
$p(\pmb{\gamma}|\hat{\pmb{\lambda}},\D)$ using the modified Langevin
algorithm can be found in the left panel of
Fig.\,\ref{ESS2018:MCMCScatterPlots:Fig} where the posterior
dependence between the vector components clearly stands out. The
reparametrization suggested in Section \ref{AsymmetricPosterior:Sec}
along the principal axes corresponding to 
the eigenvectors of the SVD decomposition of
$\widehat{\Sigma}^{\gamma\gamma}_\lambda$ yields
$\tilde{\pmb{\gamma}}$, with the scatterplot of the associated MCMC
sample
$\{
\tilde{\pmb{\gamma}}^{(m)}
=\Zeta^{-\frac{1}{2}}\mathbf{V}^\top(\pmb{\gamma}^{(m)}-\hat{\pmb{\gamma}}_\lambda)
:m=1,\ldots,M
\}$,
visible in the right panel of
Fig.\,\ref{ESS2018:MCMCScatterPlots:Fig} confirming that the posterior
dependence between the vector components of $\tilde{\pmb{\gamma}}$ is
very mild and probably negligible for most practical purposes.
%% Scatterplots
The suggested analytical approximation to
$p_{\tilde{\gamma}_s}(\tilde{\gamma}_s|\pmb{\lambda},\D)$ in
(\ref{ApproxMarginalGammaTilde:Eq2}) was evaluated and superposed to
the MCMC sample taken as a trustful proxy of the true marginal
posterior distribution of $(\tilde{\gamma}_s|\pmb{\lambda},\D)$, see
Fig.\,\ref{ESS2018:GamtildePosterior:Fig}. The quality of the
analytical approximations is excellent with a noticeable left asymmetry
for $(\tilde{\gamma}_1|\pmb{\lambda},\D)$ in particular. Transformed back in the
$\gamma$-parametrization, one obtains the analytical forms
approximated by skew-normal densities in
Fig.\,\ref{ESS2018:GammaPosterior:Fig} with again a very well
anticipated distribution of the MCMC samples for
$({\gamma}_s|\pmb{\lambda},\D)$. The positive skewness is non-negligible for the
marginal posterior distribution of $\gamma_4$: that asymmetry would
not be captured by a simple Laplace approximation. It is caused by the
small proportion of respondents in the survey expressing an explicit
disagreement with the submitted statement on the freedom of gays and
lesbians. 
%% Marginal posterior for gamma.tilde
%% Marginal posterior for gamma

\section{Discussion}
In this paper, the Laplace P-spline approach has been extended to
improve the accuracy of inference in a Bayesian framework. Indeed,
when information is sparse, the posterior distribution of
non-penalized parameters may exhibit a non-negligible skewness that
can have adverse effects on inference or predictions when ignored.  The
proposed approximation to the joint posterior density in
(\ref{JointPosteriorApprox:Eq}) takes a simple form that can be used
in a much faster way than MCMC to make predictions or inference on
functions of the model parameters.

An approximation to the marginal posterior distribution of the penalty
parameters $\pmb{\lambda}$ keeps playing an important role in the
procedure. Point estimates for $\pmb{\lambda}$ can be derived from it
with a subsequent empirical Bayes approach
\citep{CarlinLouis:2000} to handle these hyperparameters, see Section
\ref{PropOddsModel:Sec} for an 
illustration. Alternatively, the uncertainty in the selection of
$\pmb{\lambda}$ could be accounted for by marginalizing over
it with a Monte Carlo or a grid-based integration in
$p(\pmb{\xi}|\D)=\int_\lambda p(\pmb{\xi}|\pmb{\lambda},\D)
\,p(\pmb{\lambda}|\D) \,d\pmb{\lambda}.$ However, in the context of
generalized additive models \citep{gressani2021laplace} and
nonparametric double additive nonparametric location-scale models
\citep{Lambert2021fast},
simulation studies suggest that coverages of credible intervals
resulting from an empirical Bayes approach for model parameter
estimation are already close to their nominal values, even with
moderate sample sizes.

The proposed methodology diverges from the proposal made by
\citet{rueinla2009} and underlying INLA where the size of the latent vector
$\pmb{\xi}$ increases with sample size and where the marginal
distribution of the scalar components of $\pmb{\xi}$ are the research
focus. The joint distribution of $(\pmb{\xi}|\pmb{\lambda},\D)$ with
asymmetric forms for the non-penalized components is here available for
the whole vector $\pmb{\xi}$ and completed by an approximation to the
marginal posterior distribution of the hyperparameters $\pmb{\lambda}$.

\section*{Acknowledgements}
Philippe Lambert acknowledges the support of the ARC project IMAL
(grant 20/25-107) financed by the Wallonia-Brussels Federation and
granted by the Acad\'emie Universitaire Louvain.

\appendix
\section*{Appendices}

\section{Gradient and Hessian in the PO model}\label{Appendix:A}
Consider the proportional odds model defined in Section
\ref{PropOddsModel:Sec} and the notations therein.
Let $v_{ir}=F_{ir}(1-F_{ir})$,
$w_{ir}=(1 + \pi_{ir} -2 F_{ir})$,
$z_{ir}=(1-2F_{ir})v_{ir}$ for $r$ in ${\cal E}_R$ and take
$s,t\in{\cal E}_{R-1}$. One has:
\begin{align*}
  & {\partial \ell \over \partial \gamma_s}
    = \sum_i {\partial \log\pi_{iy_i} \over \partial \gamma_s}
    ~~;~~  
   {\partial \log\pi_{ir} \over \partial \gamma_s}
    = {1\over \pi_{ir}}
    (\delta_{rs}v_{ir} - \delta_{r-1,s}v_{i,r-1})\\
  & {\partial \ell \over \partial \theta_k}
    = \sum_i {\partial \log\pi_{iy_i} \over \partial \theta_k}
    ~~;~~
    {\partial \log \pi_{ir} \over \partial \theta_k}
    = x_{ik} w_{ir}
\end{align*}
and 
\begin{align*}
  & {\partial^2 \ell \over \partial \gamma_s\partial \gamma_t}
    = \sum_i {1\over \pi_{iy_i}}
    \left\{
    \delta_{y_i,s,t}z_{i,y_i} - \delta_{y_i-1,s,t}z_{i,y_i-1}
    \right\}
    -\sum_i {\partial \log\pi_{iy_i} \over \partial \gamma_s}
    {\partial \log\pi_{iy_i} \over \partial \gamma_t}~;
  \\
  & {\partial^2 \ell \over \partial^2 \theta_k\theta_\ell}
    = \sum_i x_{ik}x_{i\ell} \left(\pi_{iy_i}w_{iy_i}
    -2 \sum_{j=1}^{y_i} \pi_{ij} w_{ij} \right)~;
  \\
  & {\partial^2 \ell \over \partial\theta_k \partial\gamma_s}
    = -\sum_i x_{ik} (\delta_{y_is}v_{iy_i} + \delta_{y_i-1,s}v_{i,y_i-1})~.
\end{align*}
Therefore, given $\pmb{\lambda}$, one has
\begin{align*}
  \mathbf{U}_{\lambda}
  &= {\partial \log p(\pmb{\xi}|\pmb{\lambda},\D) \over \partial
    \pmb{\xi}}
    = {\partial \ell \over \partial \pmb{\xi}}
    -\mathbf{K}_\lambda({\pmb{\xi}}-\mathbf{e})
    ~~;~~
    \mathbf{H}_{\lambda}
  = {\partial^2 \log p(\pmb{\xi}|\pmb{\lambda},\D) \over \partial \pmb{\xi} \partial \pmb{\xi}^\top}    
  = {\partial^2 \ell \over \partial \pmb{\xi} \partial \pmb{\xi}^\top}
    -\mathbf{K}_\lambda~.
\end{align*}

% %% Bibliography
 % \newcommand{\noop}[1]{}

%\bibliography{LambertGressani}

\end{document}